\documentstyle[aas2pp4]{article}
\newcommand{\be}{\begin{equation}}
\newcommand{\ba}{\begin{eqnarray}}
\newcommand{\ee}{\end{equation}}
\newcommand{\ea}{\end{eqnarray}}

\newcommand{\etal}{{\it et al. }}
\lefthead{Nakamura  }
\righthead{Multiple Sub-Jet Model of Gamma Ray Bursts and Possible 
Interpretation for X-ray Precursor and Post-cursor}
\begin{document}
\title{Multiple Sub-Jet Model of Gamma Ray Bursts and Possible Origin
  of X-ray Precursors and Post-cursors }
\author{Takashi Nakamura}
\affil{Yukawa Institute for Theoretical Physics, Kyoto University,
Kyoto 606}
\received{2000 Jan 31 }
\accepted{2000 Mar 30}
\begin{abstract}
We assume that internal shocks of Gamma Ray Bursts (GRBs) consist of
multiple sub-jets 
with a collimation half-angle $\sim {\rm several }\times\gamma_i^{-1}$, where $\gamma_i$
is the  Lorenz factor of each sub-jet. If by chance a sub-jet is first emitted
off-axis from the line of sight, the observed peak energy can be in the X-ray region. 
Next if  by chance  a subsequent sub-jet is emitted  along the line of sight, 
 then the peak energy will be in the gamma ray region and   the gamma
ray  may arrive after the X-ray precursor from the former sub-jet
depending on parameters.  
This model predicts a new class of GRBs with  extremely weak and short  gamma ray emission but 
  X-ray precursor and/or post-cursor as well as an  afterglow.
(ApJ  Letters; Received: 2000 Jan 31	Accepted:  2000 Mar 30 )
\end{abstract}


\section{Introduction}

Recently evidences for collimation of GRBs in the afterglow  
came out for  GRB990123 (\cite{kul99,fur99,gal99,cas99}; see also
\cite{dai99} for non-jet interpretation ) and GRB990510 (\cite{hari99,sta99}).
The rapid decline rate of GRB980519 (\cite{halp99}) also suggests
the collimation of GRBs. From the fact that the power index 
of the decline rate of the afterglow 
changed at $1 \sim$ 2 days for  GRB990123 and GRB990510,  the  collimation 
half-angle $\Delta \theta_a$ of the afterglow is estimated 
as $\sim 0.1$ (\cite{rhoads97,kul99,hari99}), 
where $_a$ stands for afterglow.  
The unusually slowly declining X-ray afterglow of GRB980425 ($\propto
T^{-0.2}$)  can also be explained \footnote{In this Letter I use $T$  to express the time at
the detector.  For simplicity I  neglect the cosmological time
dilatation to avoid confusions.}  if the afterglow is beamed and 
the angle between the jet and the line of sight is $\sim 30^{\circ}$
( \cite{naka99}).

In an  internal shock model (\cite{piran98}), 
which is one of the promising models
of GRBs, the relativistic beaming half-angle ($= \gamma_i^{-1}<
0.01$) with $ \gamma_i$ being the initial gamma factor of the
internal shock is
 smaller than the suggested collimation half-angle
of the afterglow ($\Delta \theta_a \sim 0.1$).
  This means that it is possible to assume that
the collimation half-angle of the internal shock  is much smaller than the collimation
half-angle of the afterglow. We here assume that the beam of internal shocks consist
of such sub-jets with the collimation half-angle $\Delta \theta_{\gamma} <  \Delta \theta_a$.
This kind of a model has already been considered to interpret no correlation
 between the isotropic luminosity of gamma-rays and that of the
afterglow(\cite{kumar99}) as well as the diversity of the isotropic luminosity for GRBs with known red-shifts. 

Here I like to point out that such a sub-jet model has  a potential ability to
interpret X-ray precursors, which were first observed by   Ginga
satellite ( \cite{mura91}).  The evidence for X-ray precursors have been
 confirmed by GRANAT/WATCH (\cite{sazonov98}) catalogue with 6  X-ray 
precursor events in 95 GRBs.  
This suggests that about $\sim$ 6\% of GRBs have  precursors.
Recently  Beppo/SAX found the X-ray precursor and prompt X-ray 
emission from  GRB980519 (\cite{zand99}).
For  GRB980519 an analysis of the evolution of the X-ray spectrum in terms of a single power-law
spectrum shows that the photon index evolved from -2.0 to -1.1 to -2.4. The onset of burst has
 such a soft spectrum that the 2-27 keV emission appears to precede the $>$ 107 keV emission by about 70s. 
 One more important fact for GRB980519 is that the
X-ray post-cursor is seen after $\sim$ 100s from the gamma ray emission.   

\cite{pac98} has once suggested that precursors of GRBs might be the evidence for
the association of GRBs with supernovae, that is, precursors might suggest baryon contaminated
environment of GRBs, while the gamma ray might  come from the clean
fire ball. Here I like to interpret  precursors in the framework 
of the standard fire ball model which is   so successful so far (\cite{piran98}). 

In this {\it Letter}, I assume that  internal shocks of Gamma Ray
Bursts (GRBs)  consist of multiple sub-jets with 
the collimation half-angle $\sim {\rm several }\times\gamma_i^{-1}$, where $\gamma_i  \simeq 100\sim 1000$
is the  Lorenz factor of each sub-jet. If by chance a sub-jet is first emitted
off-axis from the line of sight, the observed peak energy can be in the X-ray region. 
Next if  by chance  a subsequent sub-jet is emitted  along the line of sight, 
 then the peak energy will be in the gamma ray region and depending on 
 parameters, the gamma ray  can arrive after the X-ray precursor from the former sub-jet. This
 emission pattern  agrees with that of  GRBs with precursors. The event rate
 of  precursors ($\sim 6\%$) is compatible with this model since
precursors occur by chance. If the time order of the former and the
latter sub-jets is reversed by chance,  gamma-ray comes first and
X-ray comes later. This emission pattern is compatible with the
post-cursor. In  next section we will argue a simple model in this picture. 

\section{Simple Model}

 Let us consider a simple model of GRBs in the framework of an internal
shock picture.  First at $t=0$ in the laboratory frame a slow proton jet with  Lorenz factor $\gamma_s
> 100$ with the collimation half-angle $\Delta \theta_{\gamma s}$ starts from the central engine 
 toward the direction $\theta_v$ 
where $\theta_v$ is the angle between the direction to the detector and the axis of the sub-jet. 
This slow jet continues for $\Delta t_s$. At $t=\Delta t_s+ t_r$, a rapid  proton jet 
with  Lorenz factor $\gamma_r > \gamma_s $ with the collimation half-angle 
$\Delta \theta_{\gamma r}$ starts from the central engine  toward the same direction 
$\theta_v$ as the slow jet. For simplicity we consider only the case
of $\Delta \theta_{\gamma s} = \Delta \theta_{\gamma r} \equiv \Delta
\theta_{\gamma}$. Then the rapid jet catches up the slow jet at 
$t_c=2t_r\gamma_s^2S^2/(S^2-1)+\Delta t_s$ where
  $S\equiv \gamma_r/\gamma_s > 1$.
The internal shock breaks out at $t_b=t_c+2\Delta
t_s\gamma_s^2{S'}^2/({S'}^2-1)$,
where $S'\equiv \gamma_{sh}/\gamma_s > 1$ with
$\gamma_{sh}$ being the gamma factor of the internal shock in the
laboratory frame.  
Let us consider the case of $ \theta_v > \gamma_s^{-1}$. Then the
relation of the observed time $T$ to $t$ is given by
\be
T =(1-\beta \cos\theta_v)t\sim  t\theta_v^2/2,
\ee
where we set $T=0$ as the time when the photon starts from the central
engine at $t=0$ arrives at the detector.
The observed frequency $\nu$ at the detector is related to $\nu'$ in
the shock frame as
\be
\nu =\frac{\nu'}{\gamma (1-\beta \cos\theta_v)}\sim
 11.1 \nu' (\frac{\gamma_{sh}}{200})^{-1}(\frac{\theta_v}{0.03})^{-2}.
\ee 
 Since in the internal shock model the peak
energy in the comoving frame should be in the soft X-ray band(
 say $\sim$keV) , the observed peak
energy from the off-axis emission can be in X-ray band (say $\sim$10keV)
for appropriate values of $\gamma_{sh}$ and $\theta_v$.  For
$t_r\sim \Delta t_s$,   the arrival time of this off-axis
emission ( $T_c(\theta_v)$ ) is given by
\be
T_c(\theta_v)=9t_r(\frac{\gamma_s}{100})^2(\frac{\theta_v}{0.03})^2\frac{S^2}{S^2-1},
\ee
while the duration  from the catch up time to the shock break 
out  time $\Delta T(\theta_v)$ is given by
\be
\Delta T(\theta_v)=9\Delta
t_s(\frac{\gamma_s}{100})^2(\frac{\theta_v}{0.03})^2\frac{{S'}^2}{{S'}^2-1}
\ee

Now let  us assume at $t=t_0$, the central engine starts to emit
another pair of the same slow and rapid sub-jets toward the line of
site ($\theta_v=0$). Then the catch up time ($t_c^0$) and the break
out time  ($t_b^0$) become $t_c^0=t_0+t_c$ and  $t_b^0=t_0+t_b$,
respectively, while
the relation of the observed time $T$ to $t$ becomes
\be
T(\theta_v=0)=t_0+\frac{t-t_0}{2\gamma_r^2}.
\ee 
Then
\be
T_c(\theta_v=0)=t_0+t_r\frac{1}{S^2-1},
\ee
while the duration  from the catch up time to the shock break 
out time  $\Delta T(\theta_v=0)$ is given by
\be
\Delta T(\theta_v=0)=\Delta t_s\frac{1}{{S'}^2-1}. 
\ee

In this case the observed frequency $\nu$ at the detector is related
to $\nu'$ in the shock frame as
\be
\nu = 400 \nu' (\frac{\gamma_{sh}}{200}).
\ee 
This means that the observed emission will be in the gamma ray band
if $\nu'$ is in the soft X-ray band ($\sim$keV).

Now the condition for the occurrence of the precursor 
 ($T_c(\theta_v=0) > T_c(\theta_v)$)  is rewritten as
\be
\frac{t_0}{t_r} >
(9(\frac{\gamma_s}{100})^2(\frac{\theta_v}{0.03})^2-\frac{1}{S^2})
\frac{S^2}{S^2-1}.
\ee

Let us consider a specific numerical example;

Case 1
\ba
&&\gamma_s=100, \gamma_r=200, \gamma_{sh}\sim 200,
t_r=6s, \nonumber \\
&&\Delta t_s=6s, \theta_v=0.03. 
\ea  
Then
\ba
&&T_c(\theta_v)=72s, \Delta T(\theta_v)=72s, \nonumber \\
&&  T_c(\theta_v=0)=t_0+2s,
\Delta T(\theta_v=0)=2s.
\ea
If the peak energy in the shock frame is $\sim$ 0.5keV, then the
peak energy from 72s to 144s is  $\sim$ 5keV while the peak energy
from $t_0$+2s to $t_0$+4s is $\sim$ 200keV. Therefore
 if  $t_0 > 70s$ the X-ray emission starts   $(t_0 - 70 s)$
before the gamma ray emission. The duration of the X-ray precursor
is $\sim$ 70s while the duration of the gamma ray emission is  $\sim$ 2s.
The duration of the X-ray precursor in this specific example  happens to be
 comparable to those observed by Ginga for GRB900126(\cite{mura91}) and by Beppo/SAX for 
 GRB980519 (\cite{zand99}) while the duration of the gamma ray is
somewhat smaller. 

The above model is  extremely simple  but clearly demonstrates a possible
origin of precursors. In Case 1 by choosing $t_0$ much smaller than 70s we may
explain the post-cursor. For example if we choose $t_0=20$s then the
gamma ray emission starts at 22s and ends at 24s while the X-ray post-cursor emission starts at 72s.

\section{X-ray flux from the precursor and the post-cursor in a sub-jet model}
 In the previous section, we discussed only the peak energy and the
temporal structure of precursors and post-cursors. In this section
we estimate the X-ray flux from the precursor in the sub-jet model.
 The observed X-ray flux of the precursor and post-cursor 
of  GRB980519 (\cite{zand99}) is 
$\sim 5\times 10^{-9}{\rm erg~ s}^{-1}{\rm cm}^{-2}$ in 2-10 keV band.  
The average X-ray flux of the precursor of GRB900126(\cite{mura91})
is $\sim 2.5\times 10^{-9}{\rm erg~ s}^{-1}{\rm cm}^{-2}$.
We argue  that these values are compatible with  our model of
precursors and post-cursors.

For $\theta_v=0$  and $\gamma\Delta\theta_\gamma \gg 1$,
  we assume that  the observed spectrum of the gamma
ray from the sub-jet is given by
\be
 F_{\nu}\propto \frac{1}{(1+(\frac{\nu}{\nu_0})^2)^{\beta_B/2}}, 
\ee
where $\beta_B\sim 1.5$ and $\nu_0\sim 150$keV are constants. 
The above spectrum is essentially the same as  
the Band spectrum (\cite{band93}) with $\alpha=-1$ and $\beta=-\beta_B-1$, where 
$\alpha$ and $\beta$  are the spectrum parameters in the Band
spectrum. The peak frequency $\nu_{peak}$, that is,
the maximum of $\nu F_{\nu}$, is  $\sqrt{1/(\beta_B -1)}\nu_0$. 

\cite{gran98}  as well as \cite{wood99}  derived a general formula  to compute the
off-axis emission from beamed GRBs. Here we adopt their formulations
 and notations. Let us use a spherical coordinate system ${\bf r}=(r,
\theta,\phi)$  where the coordinate are measured in the lab frame; let
 the $\theta =0$ axis ($z$-axis) points to the detector and ${\bf
r}=0$ be the central engine. Let  also   $D$ be the distance
to the source and $\alpha=r\sin \theta/D$ be the angle that a
given ray makes with the normal to the detector.  Then the observed 
 flux is given by
\ba
&&F_\nu(T) = \frac{\nu D}{\gamma\beta} 
\int_0^{2\pi} d\phi\int_0^{\alpha_m} \alpha^2 d\alpha\nonumber \\
&&\times \int_{\nu\gamma(1-\beta)}^{\nu\gamma(1+\beta)}
\frac{d\nu'}{{\nu'}^2}
~\frac{j'_{\nu'}[\Omega' ,{\bf r},T+\frac{r\mu}{ c}]}
{\{1-\mu^2\}^{3/2}},\\
&&\mu = (1-\nu'/\gamma\nu)/\beta,\\
&& \gamma =1/\sqrt{1-\beta^2},
\ea
where $\alpha_m$, $T$ and $j'_{\nu'}$ are the maximum value of $\alpha$,
 the arrival time of a photon at the detector and the rest frame
emissivity measured in ${\rm erg~s^{-1}~cm^{-3}~Hz^{-1}~sr^{-1}}$,
respectively. Note here that $'$ means the physical quantity in the
rest frame.

In the case of $\theta_v=0$, we assume 
that $j'_{\nu'}$ is expressed as
\be
j'_{\nu'}\propto H(\Delta\theta_\gamma -\theta)f(\nu '),
\ee
where $ f(\nu ')$ and $H(x)$ are  a certain 
function and the Heaviside step function, respectively.
Then $F_\nu(T)$ is given by (\cite{naka99})
\be
F_\nu(T)=2\pi A \nu\int
_{\nu/2\gamma}^{\nu\gamma\Delta\theta_\gamma^2/2}
f(\nu ')\frac{d\nu '}{{\nu '}^2},
\ee
where $A$ is a factor depending on  the density, the gamma
factor, the strength of the magnetic fields and the location of  the
shock  as well as  the distance to the source.
Since  we assume that the spectrum has the form of 
Eq. (12) for $\gamma\Delta\theta_\gamma \gg 1$, $f(\nu ')$
 is given by
\be
f(\nu ')= 
\frac{1+(\beta_B+1)(\frac{2\gamma \nu'}{\nu_0})^2}
{2\gamma (1+(\frac{2\gamma \nu'}{\nu_0})^2)^{\beta_B/2+1}}.
\ee
Neglecting the term proportional to $1/(\gamma\Delta\theta_\gamma)^2$  we have
\be
F_\nu(T)=2\pi A
\frac{1}{(1+(\frac{\nu}{\nu_0})^2)^{\beta_B/2}}
\ee
Then the  peak frequency is in the gamma ray region as $\nu_{peak}\sim \nu_0\sim 150$keV and
the peak flux is given by
\be
(\nu F_\nu)_\gamma \sim 2\pi A\nu_0/2^{\beta_B/2}.
\ee

Now let us consider the off axis case ($\theta_v >
\Delta\theta_\gamma$). 
In this case $j'_{\nu'}$ is expressed as
\ba
&&j'_{\nu'}\propto H(\cos\phi -(\frac{\cos\Delta\theta_\gamma-\cos\theta_v\cos
\theta}{\sin\theta_v\sin\theta })) \nonumber \\
&&\times H(\Delta\theta_\gamma -\mid
\theta-\theta_v\mid)f(\nu '),
\ea
The flux is given by (\cite{naka99})
\be
F_\nu(T) = 2 A \nu
 \int_{\nu\gamma(1-\beta\cos(\theta_v-\Delta \theta_\gamma) )}
^{\nu\gamma(1-\beta\cos(\theta_v+\Delta \theta_\gamma))}
\phi_v f(\nu ')\frac{d\nu '}{{\nu '}^2},
\ee
where $\cos \phi_v =(\cos\Delta\theta_\gamma-\cos\theta\cos\theta_v)/\sin\theta_v\sin \theta$.

Eq.(22) is evaluated as
\be
F_\nu(T) = 8A
(\frac{\Delta\theta_\gamma}{\theta_v})^2(\frac{1}{\gamma \theta_v })^2
2\gamma f(\nu\gamma \theta_v^2/2 ).
\ee
Then the  peak frequency is in the X-ray region as $\nu_{peak}\sim \nu_0/(\gamma
\theta_v)^2\sim 4(\gamma/200)^{-2}(\theta_v/0.03)^{-2}$keV and the peak
flux is given by
\be
(\nu F_\nu)_{\rm X}\sim
8A(\frac{\Delta\theta_\gamma}{\theta_v})^2(\frac{1}{\gamma \theta_v
})^4\frac{2+\beta_B}{2^{\beta_B/2+1}}
\ee

The ratio of the X-ray flux observed from the off axis  of the
sub-jet to the  gamma 
ray flux observed along the axis  is given by
\be
\frac{(\nu F_\nu)_{\rm X}}{(\nu
F_\nu)_\gamma}=10^{-3}(1+\beta_B/2)(\frac{\Delta\theta_\gamma}{\theta_v})^2
(\frac{\gamma}{200})^{-4}(\frac{\theta_v}{0.03})^{-4}.
\ee
For the gamma ray flux of $\sim 5\times 10^{-6}{\rm erg
~s^{-1}cm^{-2}}$, we have the X-ray flux of $\sim 5\times 10^{-9}{\rm erg~ s^{-1}cm^{-2}}$.
This X-ray flux is  similar to the observed ones in the precursors of
  GRB980519 (\cite{zand99}) and  GRB900126(\cite{mura91}). Therefore
our model  is compatible with  the X-ray flux of the precursor and the post-cursor.

\section{Discussion}
Let us try to interpret the X-ray photon index evolution of
GRB980519(\cite{zand99}) stated in Introduction. In our model
X-ray precursor and post-cursor come from the off-axis emission from
sub-jets so that the spectrum is expressed by Eq. (23) which leads to
the photon index $\beta =-\beta_B-1 \sim -2.5 $. While the main gamma
ray emission comes from the on-axis emission from the sub-jet so that 
the spectrum is expressed by Eq. (19). In Eq.(19) the photon index in
the X-ray band is $-1$. Therefore our model predicts that the photon
index evolves from -2.5 to -1 to -2.5 which agree with the observations.

As we discussed in section 2, 
the duration of the emission from the sub-jet depends on 
the viewing angle. From Eqs. (4) and (7) , we have 
\be
\frac{\Delta T(\theta_v)}{\Delta T(\theta_v=0)}=
36(\frac{\gamma_s}{100})^2(\frac{\theta_v}{0.03})^2(\frac{S'}{2})^2.
\ee
 This means that the duration of the lower energy emission ($\theta_v \neq 0$)
is  longer  than that of the higher energy emission($\theta_v = 0$).
If the X-ray precursor or post-cursors occur by chance,  our model in
this {\it Letter} predicts that their duration
are  longer than the duration of the main gamma ray emission.   
This agrees with observations of precursors and post-cursors
(\cite{mura91,zand99}).

However for most cases precursors and post-cursors do not occur.
Then what does our model predict for usual GRBs? 
In our model, in general, there exists a sub-jet
whose axis makes an angle $\theta_v\neq 0$ with the normal to the detector.
This sub-jet  mainly emits X-ray delaying from the main gamma ray
emission from the sub-jet with  $\theta_v\sim 0$. The duration of
X-ray emission is longer than the  main gamma ray emission so that
our model predicts that prompt and delayed X-ray emission with longer
duration should associate with GRBs.
The emission patterns of GRB960720, 970111, 970228,970508,
971214, 980329 and 980425 qualitatively agree with this prediction
(\cite{front99})  while other
    possible explanations may exist.

What is a new prediction from our model ?
Suppose that by chance none of the sub-jets point to the line of sight.
In this case gamma ray emission is extremely weak and short or absent while
the X-ray precursors and/or post-cursors as well as X-ray, optical and
radio afterglow exist. 
If such an event is found, that will be compatible with
 our sub-jet model discussed in this {\it Letter}.  
  
\acknowledgments
 I am very grateful Murakami, Kawai and Yoshida for useful comments.
 This work was supported in part by
Grant-in-Aid of Scientific Research of the Ministry of Education,
Culture, and Sports, No.11640274 and 09NP0801.

\end{document}